\documentclass[a4paper]{jpconf}
\usepackage{graphicx,epstopdf,epsfig,amsmath,amssymb}
\begin{document}

\title{Axion-like particles: possible hints and constraints from the high-energy Universe}

\author{Pierre Brun}

\address{CEA, Irfu, Centre de Saclay, F-91191 Gif-sur-Yvette | France}

\ead{pierre.brun@cea.fr}

\begin{abstract}
The high-energy Universe is potentially a great laboratory for searching new light bosons such as axion-like particles (ALPs). Cosmic sources are indeed the scene of violent phenomena that involve strong magnetic field and/or very long baselines, where the effects of the mixing of photons with ALPs could lead to observable effects. Two examples are archetypal of this fact, that are the Universe opacity to gamma-rays and the imprints of astrophysical magnetic turbulence in the energy spectra of high-energy sources. In the first case, hints for the existence of ALPs can be proposed whereas the second one is used to put constraints on the ALP mass and coupling to photons.
\end{abstract}

\section{Motivations for axion-like particle searches}\vspace{.5cm}

The Standard Model of particle physics reproduces data incredibly well~\cite{Beringer:1900zz}. Some of its foundations are however not completely understood, like for example the absence of CP violation in quantum chromodynamics (QCD). The most general QCD lagrangian includes a complex phase term which --if not exactly zero-- induces CP violation. The non-observation of even a very small electric dipole moment for the neutron~\cite{Baker:2006ts} implies that this phase is smaller than $10^{-11}$. This fact looks unnatural an calls for a explanation. A possible one is given by making this phase a dynamical field which value is driven to zero by the action of its classical potential. This is made possible by the introduction of a new U(1) global symmetry which is spontaneously broken at some scale $f$ (this is the so-called Peccei-Quinn symmetry~\cite{Peccei:1977hh}). A new particle that is called the axion is then predicted as an associated pseudo Nambu-Goldstone boson~\cite{Wilczek:1977pj, Weinberg:1977ma}. In the original idea of Peccei and Quinn, $f$ was of the order of the electroweak scale (EW), inducing a mass of $\sim$100 keV for the axion, which was quickly ruled out (see~\cite{Kim:1986ax, Ringwald:2012hr} for details). Then it was proposed that $f$ was much greater than the EW scale, leading to a very light and weakly interacting axion (dubbed ``invisible axion"). Axions are experimentally searched through their coupling to photons, from the Sun~\cite{2009JCAP...02..008A, 2011PhRvL.107z1302A}, assuming they make up the galactic dark matter~\cite{Asztalos:2011bm} or with LASERs~\cite{Ehret:2010mh}. Beyond the case of the strong CP problem and axions, axion-like particles (ALPs) appear in many models of physics beyond the Standard Model such as string theory~\cite{Svrcek:2006yi, Arvanitaki:2009fg, Ringwald:2012cu} as pseudo Nambu-Goldstone bosons associated to the breaking of U(1) symmetries. The properties of the associated particles are similar to that of axions, but in general their mass and coupling to photons are not related, making the corresponding parameter space larger. For the general ALP case, the interaction term with photons is
\begin{equation}
\mathcal{L}\;=\;-\frac{1}{4}g_{\gamma a} F_{\mu\nu}\tilde{F}^{\mu\nu} a \;=\; g_{\gamma a} \vec{E}\cdot\vec{B} \; a\;\;,\label{eq:1}
\end{equation}
where $g_{\gamma a}$ is the dimensionful coupling between photons and ALPs, $F$ is the electromagnetic tensor and $a$ is the ALP field. On the right hand-side of Eq.~\ref{eq:1}, the $F\tilde{F}$ term is expressed as a scalar product of the photon electric field and the magnetic field, revealing the fact that ALPs can couple to photons in the presence of an external magnetic field.

In the present article, the use of natural (astrophysical) environments to search for ALPs is emphasized, like during the propagation of very high energy gamma-rays over cosmological distances, and the effect of astrophysical magnetic turbulence on high-energy photon source  spectra. First the conventional view of the problem of the opacity of the Universe to gamma-rays is presented, with the discussion of a possible indication for an anomalously transparent Universe. Although the possible tensions can be solved in a conventional way, they can also be released by invoking ALPs mixing with photons. Then it is shown that this observable could be used as a signature when one tries to make a discovery, but that some uncertainties prevent from using it to derive robust constraints.
It is then shown that constraints can be obtained by considering the effect of magnetic turbulence around the sources and finally some examples of constraints are given.

\section{The transparency of the Universe and ALPs}\vspace{.5cm}

\subsection{The conventional view of the Universe opacity to gamma-rays}\vspace{.5cm}

\begin{figure}[h]
\centering
\includegraphics[width=.5\textwidth]{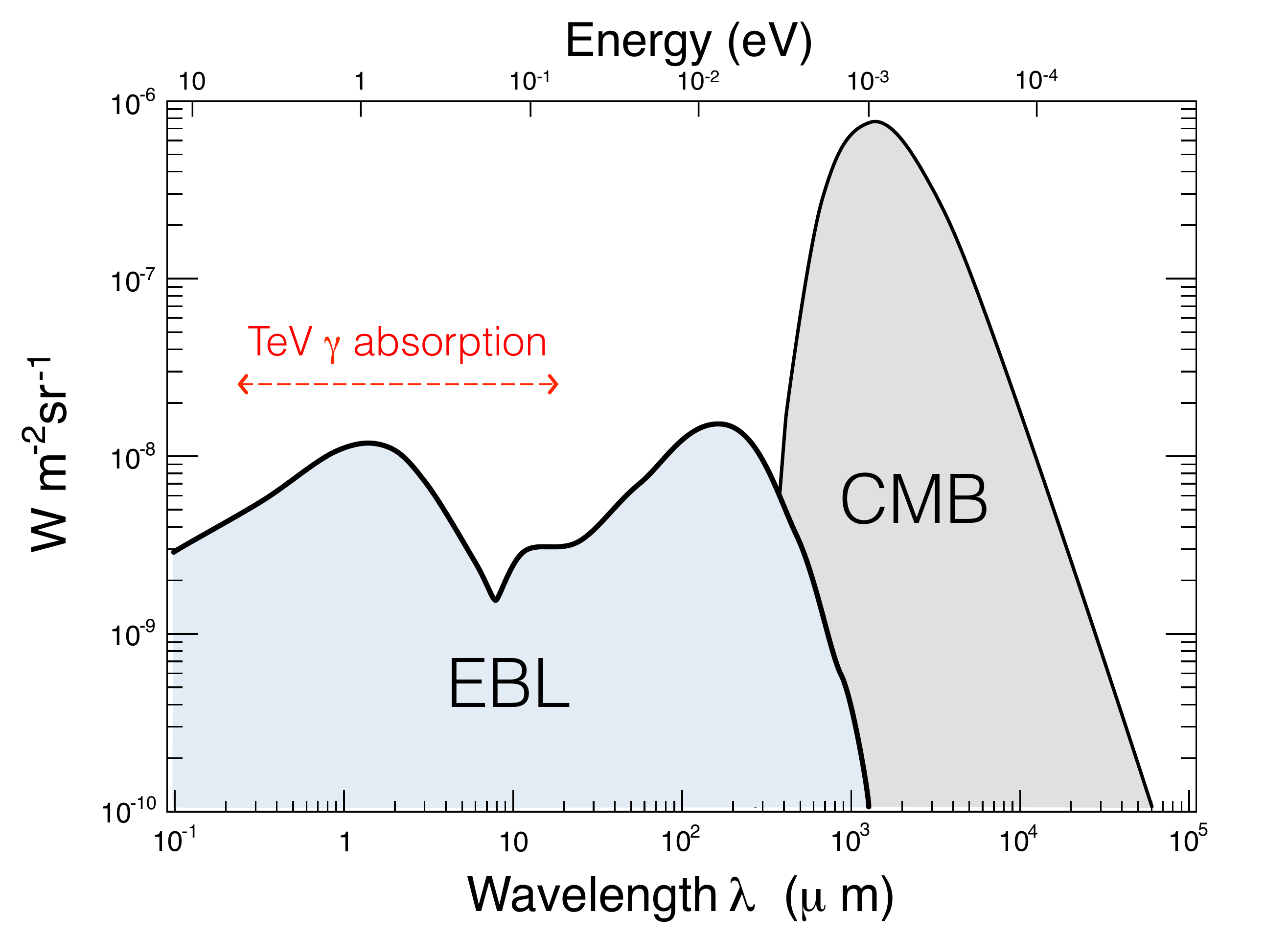}
\includegraphics[width=.4\textwidth]{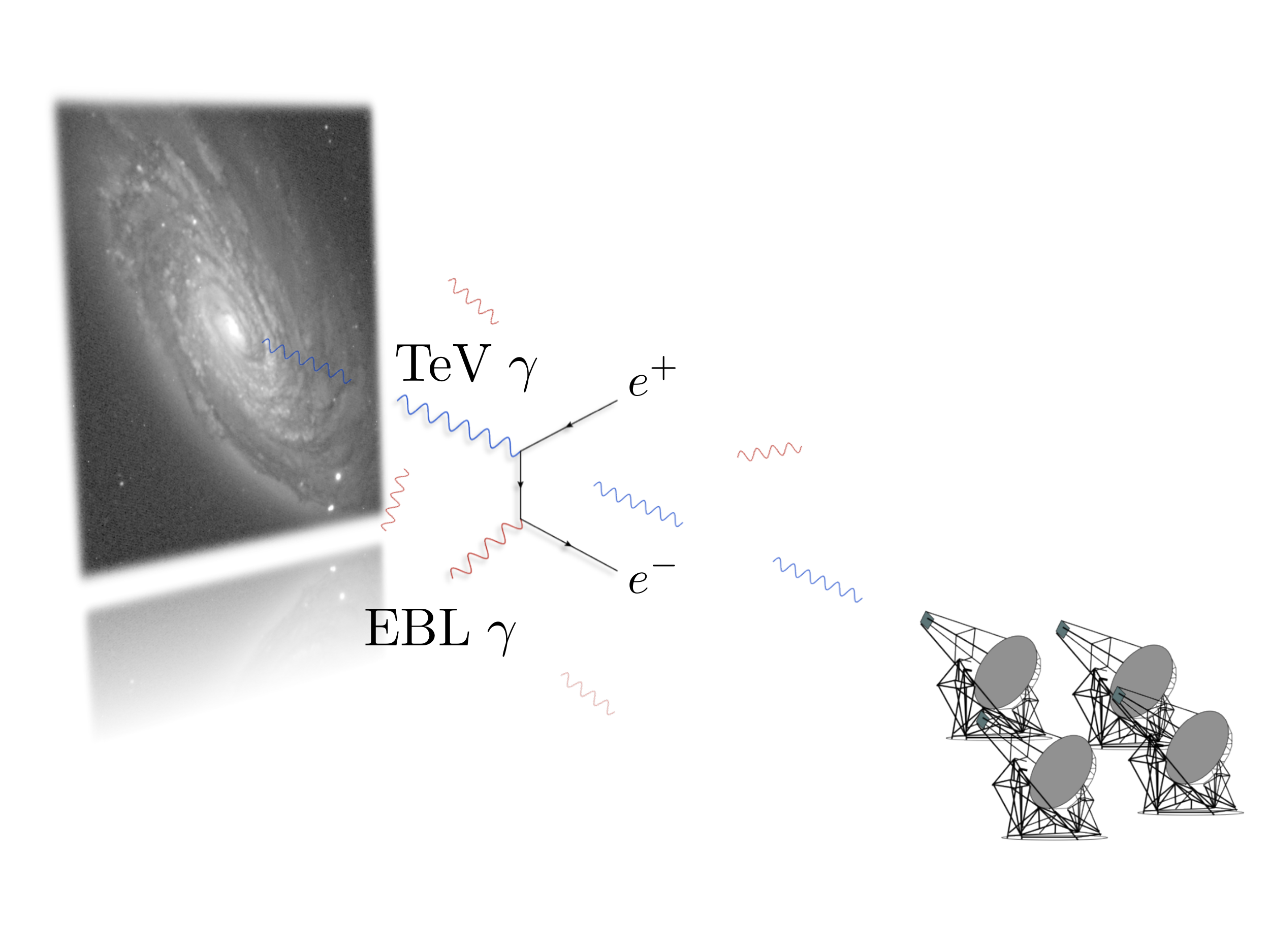}
\caption{Left: Spectral energy density of the cosmic background photons including the CMB and the EBL (inspired from~\cite{Dole:2006de}, with permission of the authors). Right: Illustration of the pair production process.\label{fig:EBL}}
\end{figure}

Very high energy photons (with $\sim$TeV energies) traveling through the intergalactic medium encounter different populations of background radiations. The most numerous type of background photons belong to the Cosmic Microwave Background (CMB) and a second population is the extragalactic background light (EBL). The latter has a double bump structure, that comes from direct starlight and emission re-processed by interstellar dust in the infrared band as sketched on Fig.~\ref{fig:EBL}. Direct measurement of the EBL is very difficult because of foregrounds and infrared radiation by the instruments.  

\begin{figure}[h]
\centering
\begin{minipage}[c]{.49\textwidth}
\centering
\includegraphics[width=\textwidth]{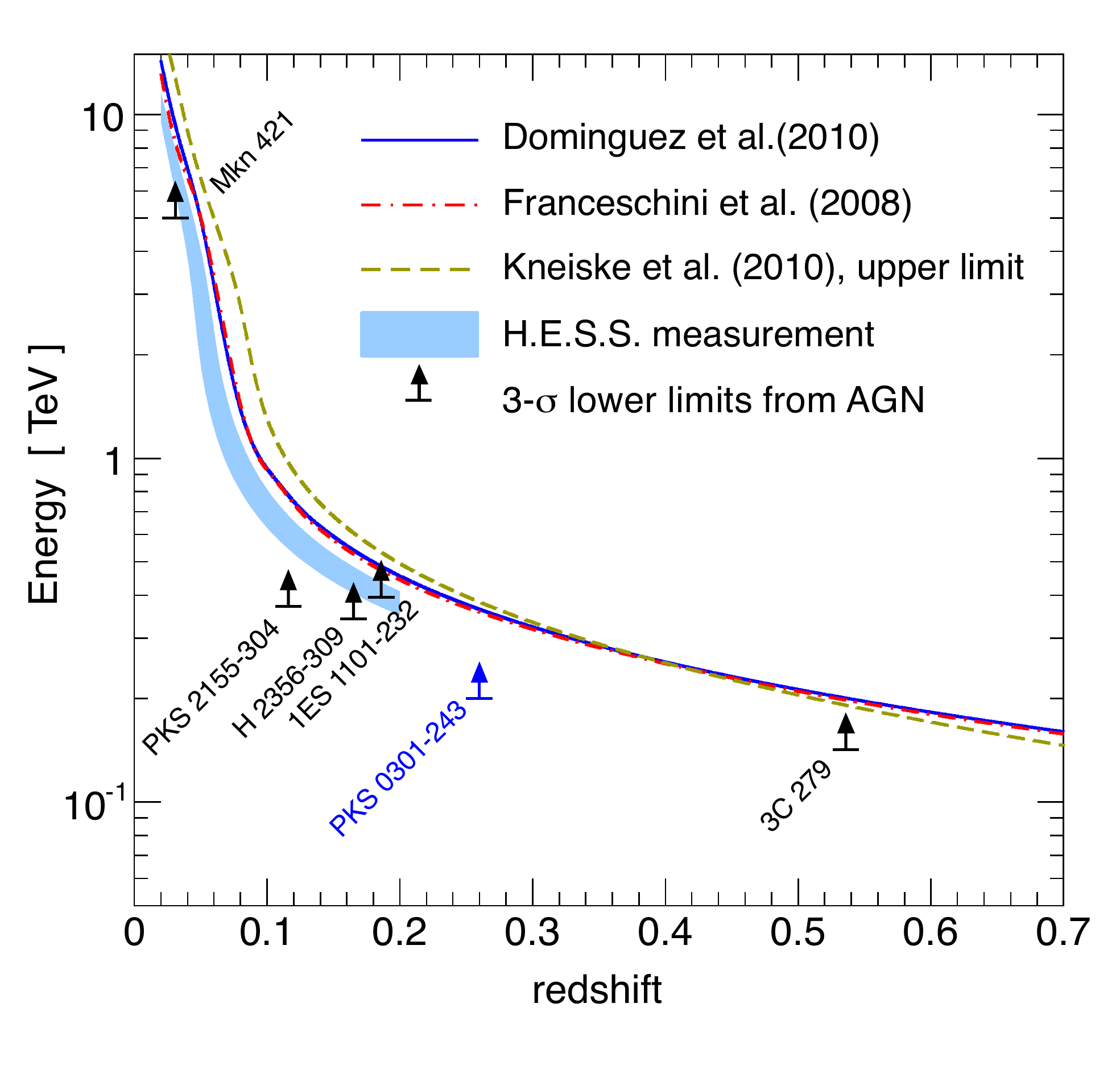}
\end{minipage}
\begin{minipage}[c]{.5\textwidth}
\centering
\caption{Energies corresponding to $\tau=1$ for different EBL models, constraints from very high energy gamma-ray astronomy and 1-$\sigma$ measurements from HESS (figure from~\cite{Wouters:2012ek, hess}).\label{fig:eblconstr}}
\end{minipage}
\end{figure}

TeV gamma-ray astronomy is sensitive to the EBL density and spectrum as it is responsible for the attenuation of extragalactic source fluxes at high energy. The reason for that is the pair production process $\gamma_{\rm TeV}\gamma_{\rm EBL}\rightarrow e^+e^-$, for which the threshold lies at TeV energies in the terrestrial frame. For instance considering $E_{\rm EBL}\sim 0.1\;\rm eV$, the threshold energy satisfying $E_{\rm th}E_{\rm EBL}>m_e^2$ (where $m_e$ is the mass of the electron) yields $E_{\rm th}\sim2.6\;\rm TeV$. In Fig.~\ref{fig:EBL} the typical range of the TeV absorption range is indicated by the horizontal arrow. Because of the pair production process, the highest energy photons have a larger optical depth. Before 2006 it was commonly admitted that is was very unlikely to detect TeV photons from sources above $z\sim 0.2$. The situation changed after HESS observations of two active galactic nuclei (AGNs) at $z=0.186$ and $z=0.165$. As reported in~\cite{Aharonian:2005gh}, when unfolded from the EBL effect, the intrinsic spectra of these sources were found too hard and in tension with the source models. This was the first indication for a Universe slightly more transparent than expected at high energy. Later, AGNs were observed at redshifts as high as 0.536 by MAGIC~\cite{Aliu:2008ay} and possibly 0.61 by HESS~\cite{Becherini:2012ei}. The Universe is indeed more transparent to gamma-rays that expected. That puzzle has conventional solutions, it could be for instance that spectra are actually harder, this can be realized for instance including hadronic components in the AGN jets or in relativistic shock acceleration models. The tension can be removed as well with a revision of the EBL models, mainly with a lower density as for instance in the model of~\cite{Franceschini:2008tp} which is compatible with all observations. TeV observations are now even used to provide not only upper limits on the EBL density but actual measurements~\cite{Abramowski:2012ry}. The current situation is illustrated in Fig.~\ref{fig:eblconstr}, extracted from~\cite{Wouters:2012ek, hess}. It represents the energy above which the absorption becomes significant (defined by a optical depth $\tau(E)=1$) as a function of the redshift of the source. The lines correspond to models or lower limits for the EBL density and redshift evolution from~\cite{Franceschini:2008tp, Dominguez:2010bv, Kneiske:2010pt}. Constraints from the spectral indices of different sources are shown as arrows, and the HESS measurement corresponds to the blue band. One source seems to be in tension with the measurement. However the methods that lead to the constraint and the measurements are different as the constraints rely on spectral slope measurements and the measurement comes from the observation of features in the spectra that can be related to the EBL spectral density. A more unified approach might be necessary to get a definite answer on how strong the tension is. Note also that the Fermi collaboration did the same measurement at higher redshifts and found a good compatibility with EBL models~\cite{:2012gd}.

Some studies however still claim for an anomaly, even with the lower EBL limits from~\cite{Kneiske:2010pt}. It is the case in~\cite{Horns:2012fx}, where the authors claim for an anomaly, with the caveat that their claim requires to leave out some error bars.

\subsection{How ALPs enter the game}\vspace{.5cm}

\begin{figure}[h]
\centering
\begin{minipage}[c]{.48\textwidth}
\centering
\includegraphics[width=\textwidth]{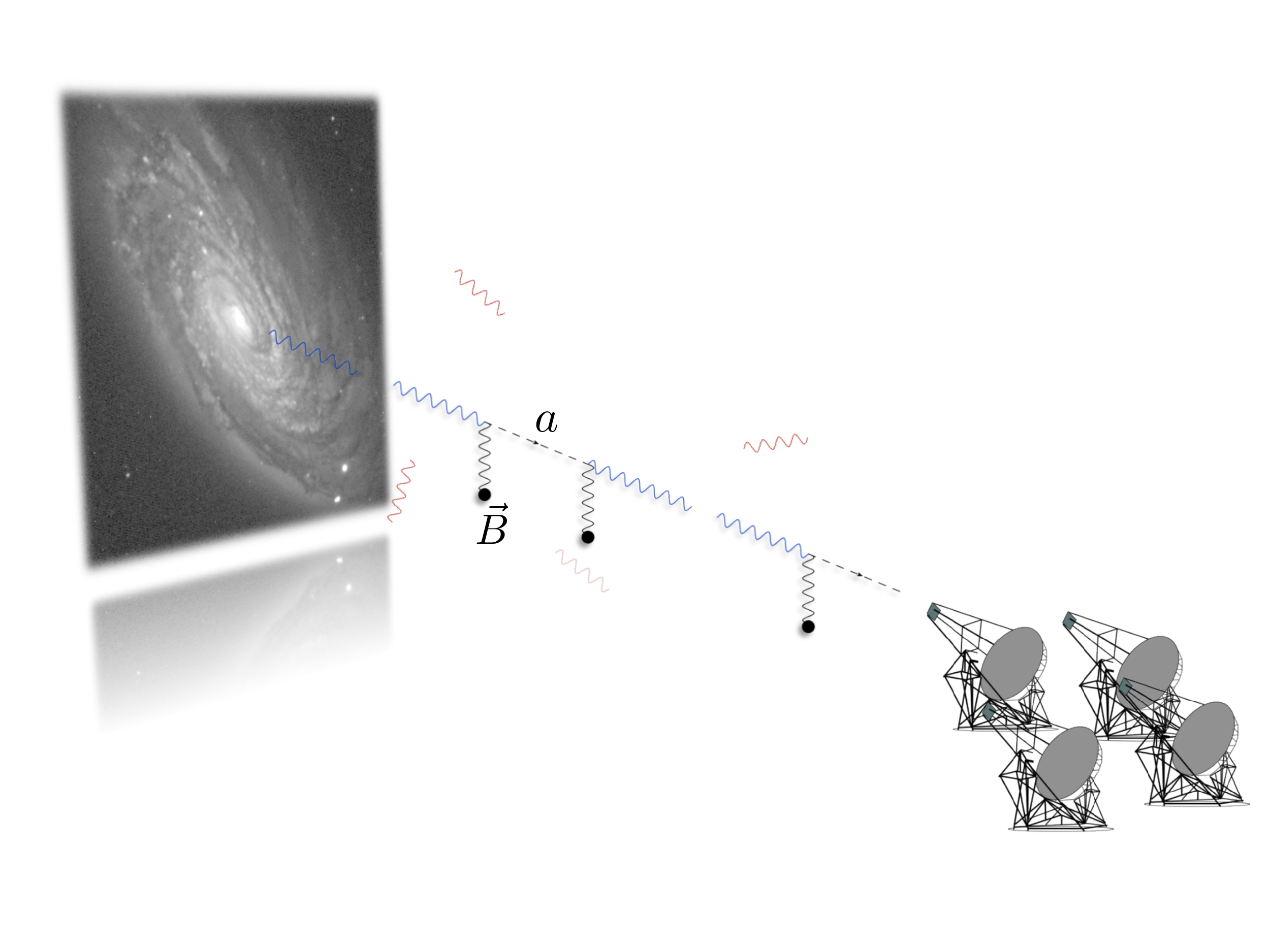}
\caption{Illustration of the photon/ALP oscillations in a magnetic field\label{fig:ALPs}}
\end{minipage}
\hspace{0.01\textwidth}
\begin{minipage}[c]{.48\textwidth}
\centering
\includegraphics[width=\textwidth]{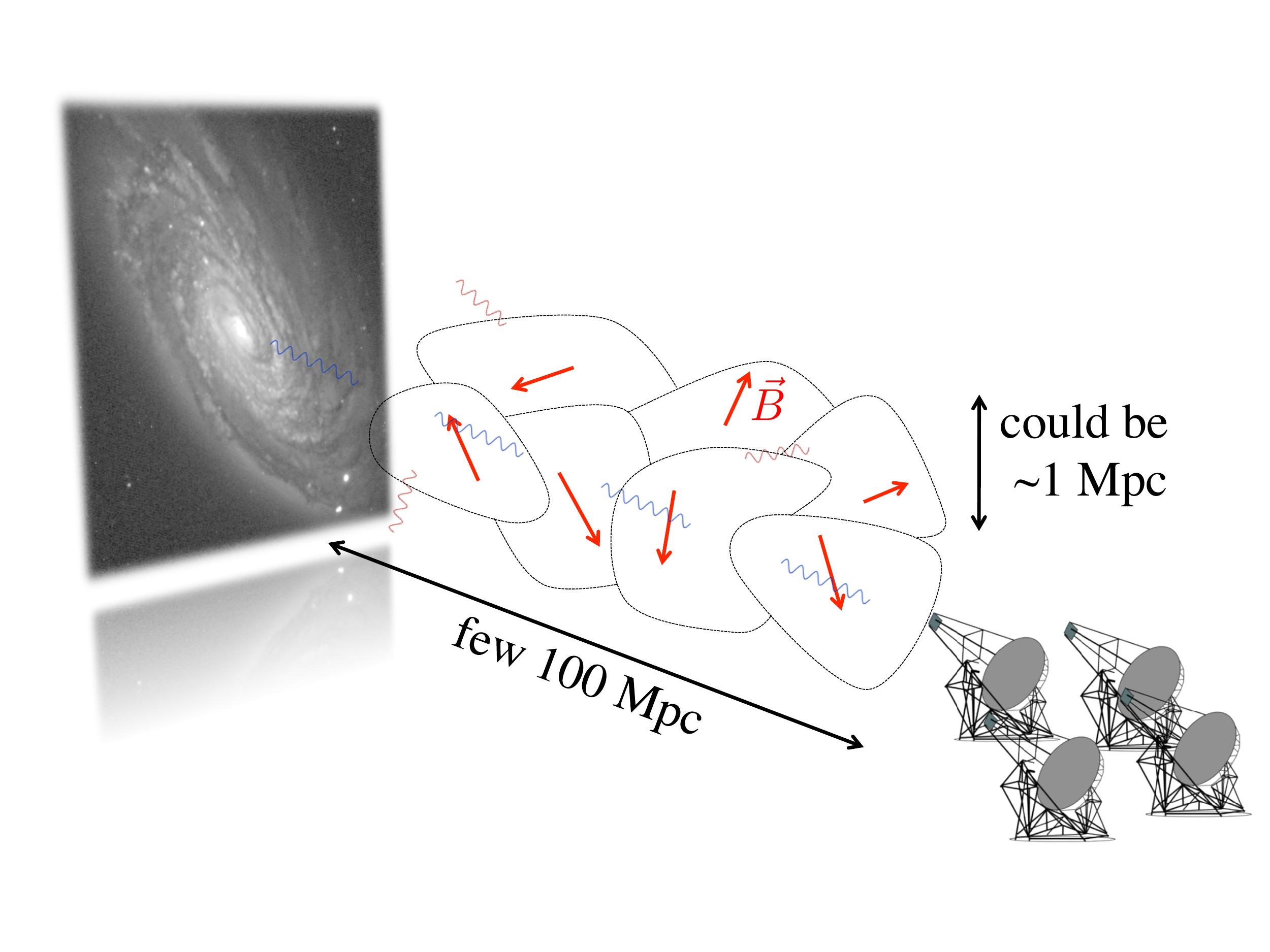}
\caption{Illustration of the modeling of the extragalactic magnetic field\label{fig:mag}}
\end{minipage}
\end{figure}

The lack of opacity of the Universe to gamma-rays gave rise to the idea that ALPs could be responsible for this effect. The basic idea is that if mixing between ALPs and photons occur, the beam could travel in the form of ALPs on a significant fraction of way, not producing pairs, as sketched on Fig.~\ref{fig:ALPs}. If ALPs are converted back to photons before observations, this could lead to a more transparent Universe.

To get an idea of the relevant masses and couplings for the ALPs that are invoked here, let us consider the most simple formalism for describing the photon/ALP mixing. The system propagation is described by a Schr\"odinger-like equation:
\begin{equation}
(E-i\partial_z-\mathcal{M})
\left(\begin{array}{c} A\\a \end{array}\right )
\;=\;0
\;\;
\text{with}
\;\;
\mathcal{M}\;=\;
\left(\begin{array}{cc}
-i\frac{\tau}{2z} & \Delta_B \\
\Delta_B  & \Delta_a
\end{array}\right )
\;\;,
\end{equation}
where $\Delta_B=g_{\gamma a}B_t/2$ describes the photon/ALP coupling ($B_t$ is the transverse projection of the magnetic field), $\tau$ is the optical depth related to EBL absorption and $\Delta_a=-m_a^2/2E$ accounts for the ALP mass. The fact that the imaginary coefficient only applies to the photon part of the wavefunction leads to the change in the overall transparency. In the case of no absorption, the mixing matrix is diagonalized with a rotation angle $\theta$ such that $\tan 2\theta = -2\Delta_B/\Delta_a$. The resolution of the propagation equation in the propagation state basis leads to the probability of transition
\begin{equation}
P_{\gamma\rightarrow a}\;=\;\frac{1}{2}\frac{1}{1+\left (E_c/E \right )^2}\sin^2\left (\frac{g_{\gamma a} B_t \;z}{2} \sqrt{1+\left (\frac{E_c}{E}\right )^2} \right ),\;\;\text{with} \;\; E_c=\frac{m_a^2}{2 g_{\gamma a} B_t}\;\;.
\end{equation}
The overall 1/2 coefficient in $P_{\gamma \rightarrow a}$ accounts for the two polarizations of the photon. A critical energy $E_c$ appears, that defines the energy scale at which strong mixing occurs. From the expression of $E_c$, with cosmological magnetic fields $B\sim 1\;\rm nG$, an ALP mass $m_a\sim \rm neV$ and a coupling $g_{\gamma a}\sim 10^{-11}\;\rm GeV^{-1}$, the critical energy lies at the TeV scale. It follows that the type of ALPs that are concerned by the so-called transparency hint will fall in a region of low masses and with couplings larger than those of the corresponding axions. 

The full treatment of the transparency problem in the presence of ALPs requires a $3\times 3$ mixing matrix to account for the two polarization states for the photon, and a description of the magnetic field on the path from the source to the observatory. The extragalactic magnetic field is usually described as a patches of coherent domains of 1 Mpc size. The magnetic field strength is the same in all domains but from one domain to the next its orientation changes in a random way (see the sketch of Fig.~\ref{fig:mag}). It can be shown (see~\cite{Grossman:2002by}) that for random orientations and a large number $N$ of domains, the transition probability is reduced to
\begin{equation}
P_{\gamma\rightarrow a}\;=\;\frac{1}{3}\left ( 1- \exp\left( -3NP_0 \right ) \right )\;\;,
\end{equation}
where $P_0$ is the transition probability in one domain. From this expression one would expect to have a $1/3$ drop in the energy spectrum above $E_c$ in the limit $NP_0\gg1$, and a flux that is boosted at high energy (typically above the pair-production related cutoff) as described in~\cite{SanchezConde:2009wu}.

\begin{figure}[h]
\centering
\includegraphics[width=.8\textwidth]{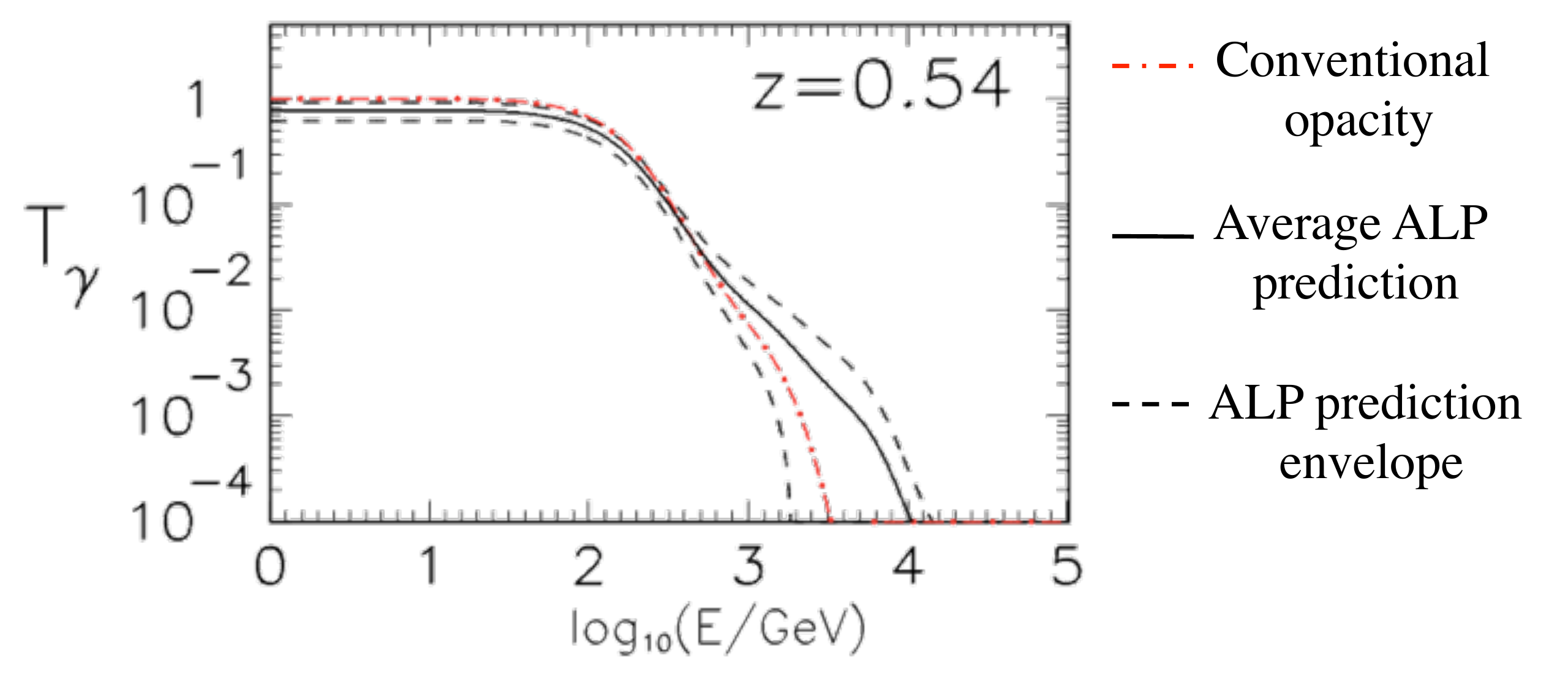}
\caption{Transmission of photons with and without ALPs, the ALP case is ploted with the envelope corresponding to the variance on the prediction of the transparency effect (figure from~\cite{Mirizzi:2009aj}, with permission of the authors).\label{fig:stoch}}
\end{figure}

At least two facts lead to revise the above statements. First, in practice the limit $NP_0\gg1$ is hardly realized. Second, due to the unknown nature of the magnetic field configuration, the prediction on the transmission has an intrinsic variance. Indeed it can happen that the ALPs do not convert back into photons before reaching the Earth, leading in that case to an even more opaque Universe. This is nicely illustrated in Fig.~\ref{fig:stoch} extracted from~\cite{Mirizzi:2009aj}. Here the red dot-dashed line corresponds to the conventional opacity in the absence of ALPs and the solid black line is the average prediction with ALPs. It appears that the average transparency is indeed higher than the conventional case at high energy. However, the associated uncertainty on the prediction, in other words the variance related to the randomness of the magnetic field is such that the envelope includes the conventional case. Because of that fact, if observed without ambiguity in the future, such an effect might be seen as an indication for ALP detection but could hardly serve as a firm argument for discovery.

\begin{figure}[h]
\centering
\begin{minipage}[c]{.64\textwidth}
\centering
\includegraphics[width=.9\textwidth]{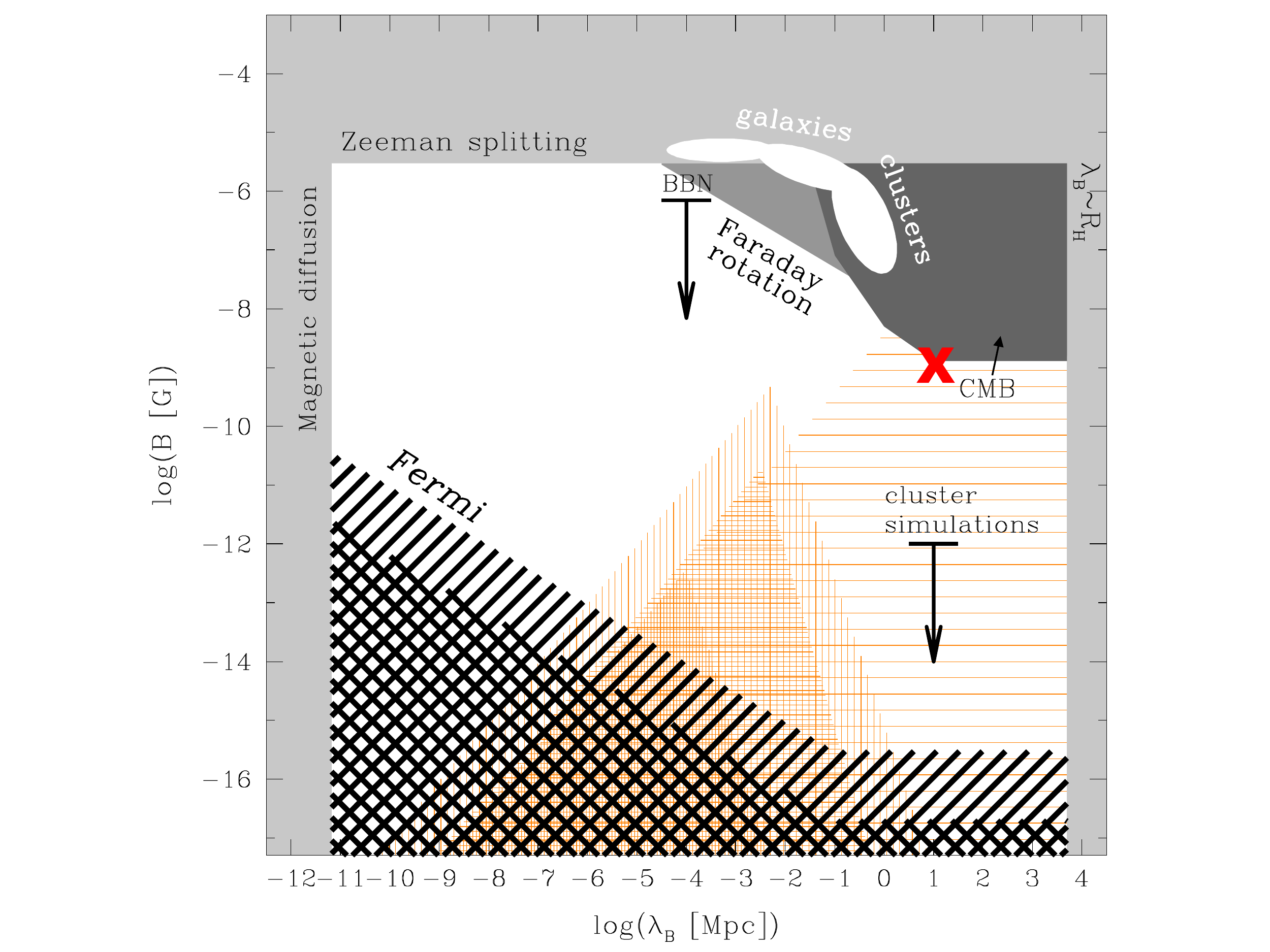}
\end{minipage}
\begin{minipage}[c]{.35\textwidth}
\centering
\caption{Constraints on large scale magnetic fields and prediction of the models. The red cross corresponds to the parameters used for the ALP solution to the transparency ``hint" (figure adapted from~\cite{Neronov:1900zz}, with permission of the authors).\label{fig:magfield}}
\end{minipage}
\end{figure}

Another limitation comes from the use of a very optimistic value for the extragalactic magnetic field. For the ALP effect to significantly affect the opacity, magnetic fields of nG strength with Mpc coherence length have to be present in the intergalactic medium. It is actually possible to generate such magnetic fields from inflation or QCD phase transition for instance, but the required strength is very close to current upper limits. This is illustrated in Fig.~\ref{fig:magfield} (from~\cite{Neronov:1900zz}), where the different observational constraints on large scale magnetic field appear together with predictions from models (orange thin lines). There the red cross corresponds to the typical parameters used in the ALP analyses. It lies in a region that can be seen as fine-tuned given the size of the still open parameter space. At the moment it seems invoking such a strong magnetic field would be acceptable if the tension in the TeV observations was stronger.

A clever way to avoid using intergalactic magnetic fields is to remark that if the source is magnetized or embedded in a cluster, then the mixing to ALPs could occur essentially around the source. Then, the magnetic field of the Milky-Way can serve as a target magnetic field to convert back the ALPs into photons. In that case as well, a strong boost can be expected at high energy, as first proposed in~\cite{Simet:2007sa} and then again in~\cite{Horns:2012kw}. This effect is used in~\cite{Meyer:2013pny} to estimate possible lower limits on the $g_{\gamma a}$ coupling, but again assuming the tension is real between observations and models for the transparency. Finally it appears that the transparency observable could be used in the future for indication or discovery, if a clear tension was observed. To do so, one might wait for the next generation of gamma-ray telescopes such as CTA to have a significantly larger sample of sources.

The problem of having only a few sources can be circumvented by using an energy band for which detections are numerous. This has been proposed in~\cite{Burrage:2009mj}, where the authors remark that if the strong mixing regime is realized, the statistical properties of the observed fluxes from X-ray sources could display features distinctive of ALP effects. In that case, sources would be seen with fluxes reduced by a factor of 1/3 on average. Of course having no access to the absolute intrinsic fluxes, this overall factor is not observable. However because of the random nature of the mixing process in astrophysical magnetic fields, the first and second momentum distribution should have different shapes compared to the conventional case. In~\cite{Burrage:2009mj} the authors claim the observation of anomalous features in the momenta distributions. That result has however shown to be questionable in~\cite{Pettinari:2010ay} where the effect is claimed to be caused by outliers. Because in that case the detection would rely on shapes of distributions, it is difficult to infer a constraint without a deeper analysis and this effect is again used to propose a hint. Nevertheless, it illustrates one possible use of the stochastic nature of the mixing in astrophysical environments, which is no more a limitation but becomes a tool for identifying possible ALP effects. In the following, it is shown that a careful study of this randomness can lead to observable effects that are used to set constraints on the ALP parameters.

\section{Constraints on ALP parameters from observations of the high-energy sky}\vspace{.5cm}

\subsection{Effect of the magnetic turbulence}\vspace{.5cm}

One peculiar effect of photon/ALP mixing is the fact that the magnetic field turbulence can directly imprint features in the energy spectra from high-energy sources. The exact spectral shape one gets at the end is unpredictable, but as shown in~\cite{Wouters:2012qd} the statistical properties of the induced irregularities are a prediction of the ALP model. The authors of~\cite{Ostman:2004eh} and~\cite{Mirizzi:2009aj} already noticed that in principle the observed spectra should be very irregular in case of strong photon/ALP mixing, without considering the use of the irregularity as an observable.

\begin{figure}[h]
\centering
\begin{minipage}[c]{.48\textwidth}
\centering
\includegraphics[width=\textwidth]{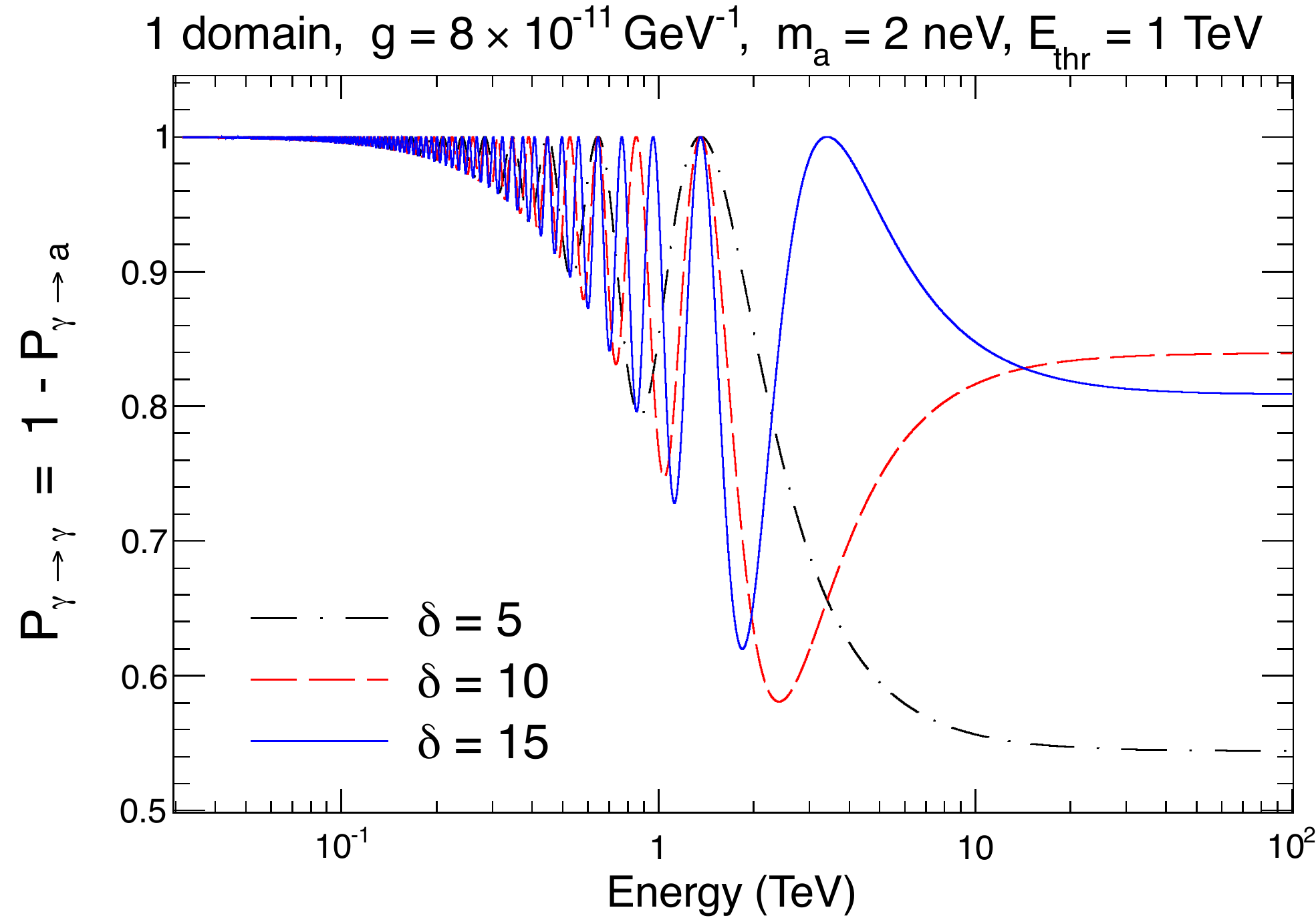}
\caption{Spectral oscillation patterns in domains with coherent magnetic field and different ALP parameters (figure from~\cite{Wouters:2012qd}).\label{fig:osc}}
\end{minipage}
\hspace{.01\textwidth}
\begin{minipage}[c]{.48\textwidth}
\centering
\includegraphics[width=\textwidth]{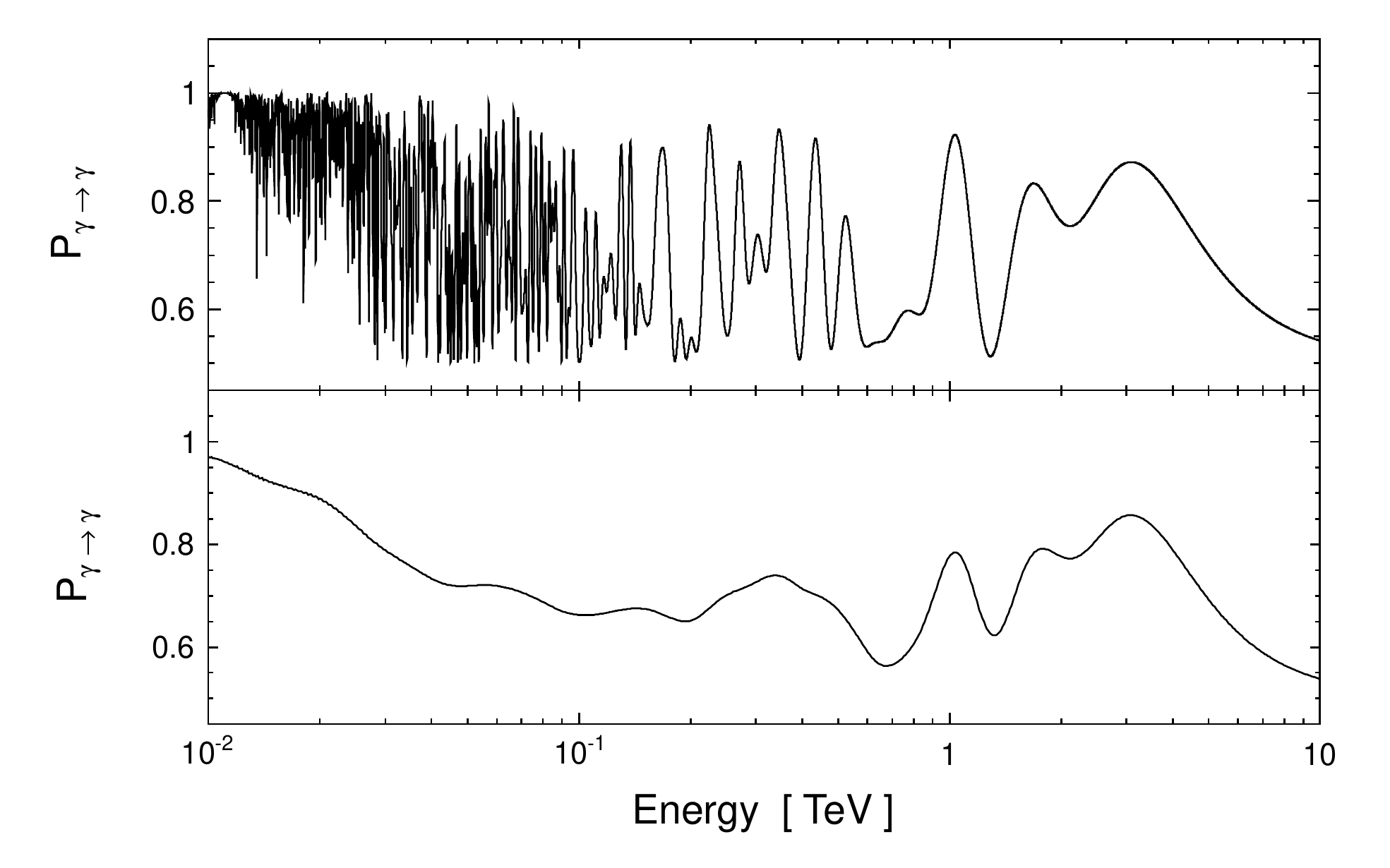}
\caption{Example of ALP induced irregularity in the TeV range (top panel: Raw signal, bottom panel: Signal smeared with HESS resolution, figure from~\cite{Wouters:2013iya, hess2}).\label{fig:irr}}
\end{minipage}
\end{figure}

To account quantitatively for the irregularity, the 2 polarizations of the photon must be considered, so that the evolution of the system after $n$ domains is given by
\begin{equation}
|\psi_n\rangle\;=\;\prod_k \left ( P_k^{-1}\; \exp \left [ -i \left (E + \mathcal{M}^\star_k \right ) s_k\right ]\; P_k\right )\; |\psi_0\rangle\;\;,\;\;\;\;\text{with}\;\;\mathcal{M}^\star_k=P_k \mathcal{M}_{k} P_k^{-1}\;\;
\end{equation}
and
\begin{equation}
\mathcal{M}_k\;\;=\;\;\left(\begin{array}{ccc} 
-\frac{m_\gamma^2}{2E}-i\frac{\tau}{2z} 	& 0 									& \frac{1}{2} g_{\gamma a} B_t^{(k)}\cos\phi^{(k)} \\ \\
0 								& -\frac{m_\gamma^2}{2E}-i\frac{\tau}{2z}  	& \frac{1}{2} g_{\gamma a} B _t^{(k)} \sin\phi^{(k)}\\  \\
\frac{1}{2} g_{\gamma a} B_t^{(k)} \cos\phi^{(k)}			& \frac{1}{2} g_{\gamma a} B_t^{(k)}\sin\phi^{(k)} 			& \frac{-m_{\rm a}^2}{2E} 
\end{array}\right)\;\;. \label{eq:33mat}
\end{equation}
$k$ stands for the $k^{\rm th}$ domain, of size $s_k$, $P_k$ is the rotation matrix between the interaction eigenstates and the propagation eigenstates and the matrix $\mathcal{M}_k$ describes the mixing. The indexes $k$ are there to recall that from one magnetic domain to the next, the corresponding parameters change due to the different orientations of the magnetic field ($B_t$ is the projection of the magnetic field on the polarization plane and $\phi$ is the angle that projection makes with one of the two photon polarization). $m_\gamma = 4\pi\alpha n_e/m_e$ is the effective mass of the photon propagating in a plasma with electron density $n_e$. Examples of spectral oscillation patterns in one domain are given in Fig.~\ref{fig:osc} for different values of $\delta=g_{\gamma a} B_t s/2$, $s$ being the size of the coherent domain. When several domains are considered, the spectrum ends up being very irregular as shown in Fig.~\ref{fig:irr} in the case of an unpolarized beam. For that example, an extragalactic source is considered and the magnetic field is typical of that of a galaxy cluster. The top panel of Fig.~\ref{fig:irr} is the raw signal and the bottom panel is the same signal smoothed by the energy resolution of HESS ($\sim$15\%). In that case the critical energy is of order 1 TeV and the effective photon mass is negligible.

\begin{figure}[h]
\centering
\begin{minipage}[c]{.4\textwidth}
\centering
\includegraphics[width=.9\textwidth]{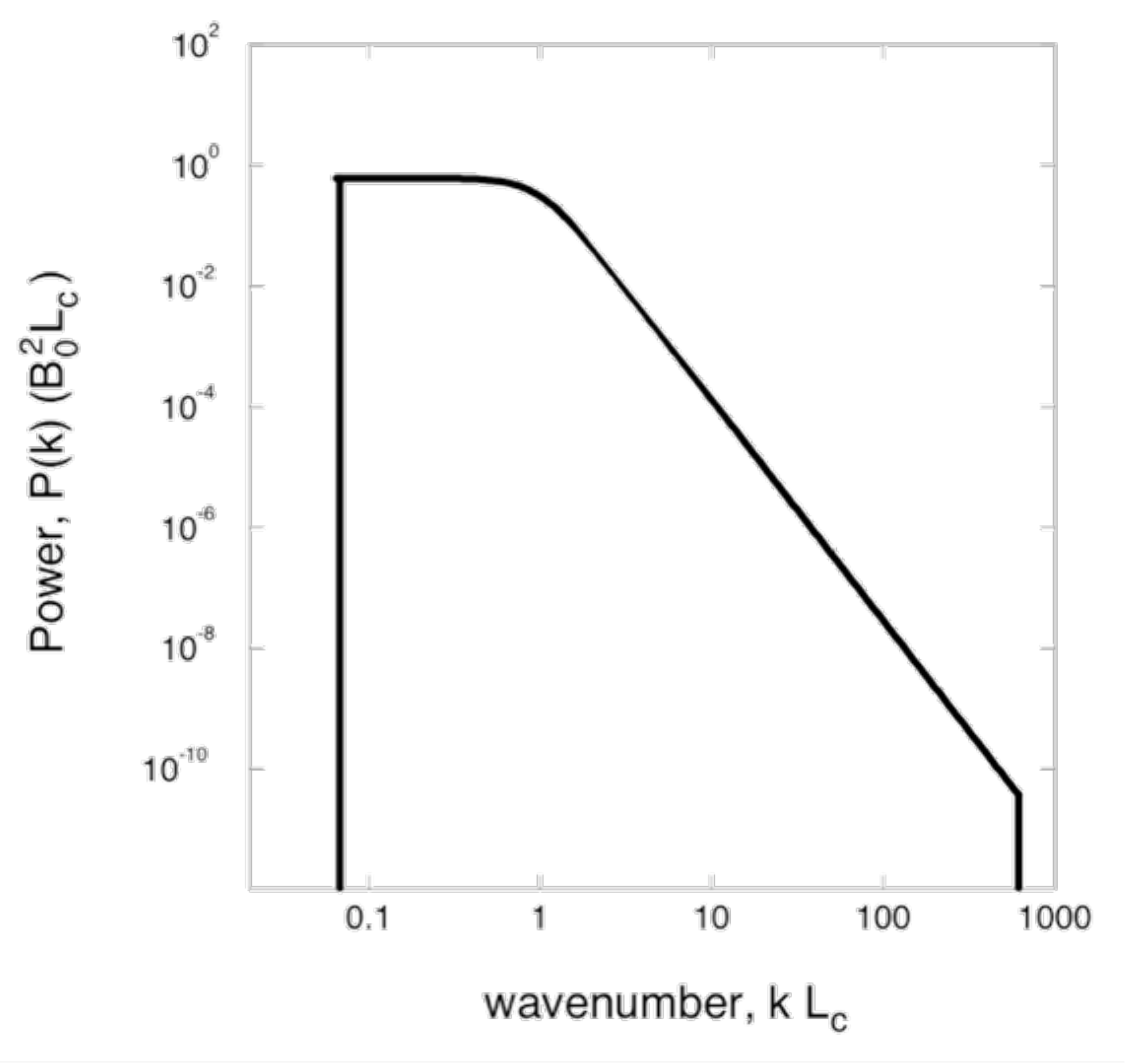}

%
\end{minipage}
\hspace{.09\textwidth}
\begin{minipage}[c]{.49\textwidth}
\centering
\caption{Typical power spectrum used for modeling the magnetic turbulence in galaxy clusters.\label{fig:turb}}
\end{minipage}
\end{figure}

Whereas in the case of the extragalactic magnetic field the naive description of the turbulence might be sufficient (essentially because its properties are very poorly known), galaxy cluster magnetic fields may deserve a better treatment. The magnetic field in that case is modeled by a gaussian field with zero mean and a distribution of modes that is described by a Kolmogorov-like spectrum as in Eq.~\ref{eq:turb}:
\begin{equation}
\left (\delta B\right )^2 \propto \sigma^2\frac{k^2}{1+(kL_{\rm c})^\gamma}\label{eq:turb}\;\;.
\end{equation}
The corresponding power spectrum is modeled by a function resembling that of Fig.~\ref{fig:turb}. In galaxy clusters, the typical coherence length of the magnetic field is 10 kpc and the strength of the field is 1 to 10 $\mu$G. 

\subsection{Examples of constraints}\vspace{.5cm}

\begin{figure}[h]
\centering
\begin{minipage}[c]{.48\textwidth}
\centering
\includegraphics[width=\textwidth]{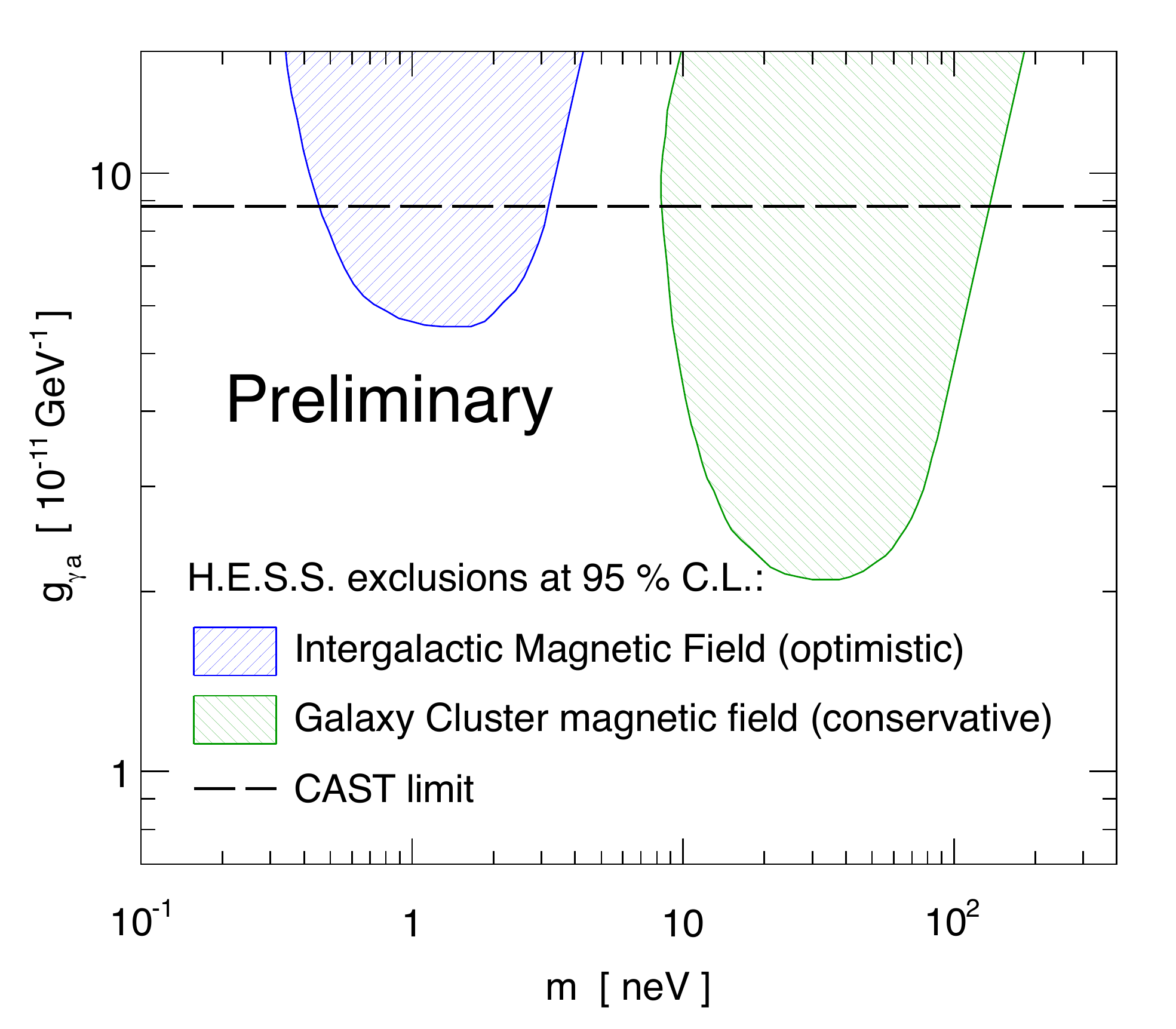}
\caption{HESS exclusion contours from the analysis of PKS 2155-304 energy spectrum (Figure from~\cite{Wouters:2013iya, hess2}).\label{fig:excl_hess}}
\end{minipage}
\hspace{.01\textwidth}
\begin{minipage}[c]{.48\textwidth}
\centering
\includegraphics[width=.9\textwidth]{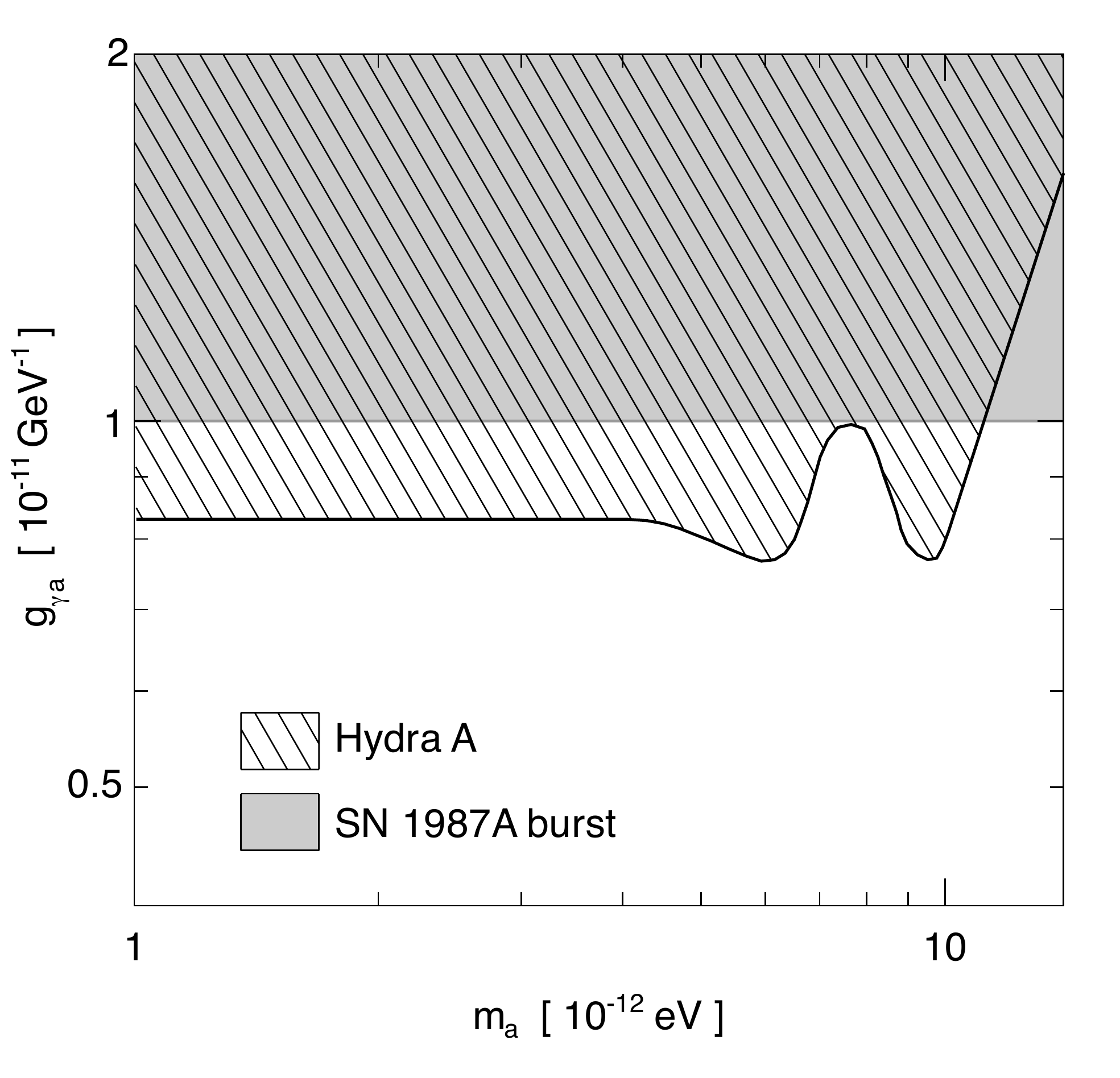}
\caption{Exclusion contours from the analysis of X-ray data from Hydra A (Figure from~\cite{Wouters:2013hua}).\label{fig:excl_x}}
\end{minipage}
\end{figure}

The first example of constraints is from the HESS analysis of PKS 2155-304~\cite{Wouters:2013iya, hess2}, which is an AGN located at $z=0.116$. For that source, both the extragalactic magnetic field and the cluster magnetic field can be considered. In the first case, as previously, one has to assume optimistic values of the magnetic field strength for the irregularity signal to be significant. A galaxy cluster is observed around the source, but no magnetic field measurements are available. So in the case of the galaxy cluster magnetic field, conservative values for the strength and the coherence length are assumed (1 $\mu$G and 10 kpc respectively). As HESS observation ranges from hundreds of GeV to a few TeV, from the expression of the critical energy $E_c$ it is straightforward to see that the typical ALP masses that are probed are of the order of $10^{-8}$ eV. In~\cite{Wouters:2013iya, hess2}, it is shown that the observed energy spectrum does not exhibit strong irregularities. Then an estimator of the irregularity is proposed and numerical simulations are used to exclude sets of parameters that lead to significantly too strong irregular behavior. This exclusion has to be done on a statistical basis as each realization of the magnetic field turbulence is different. The results of the analysis are presented in Fig.~\ref{fig:excl_hess}. The method allows to improve the CAST limits in a limited energy range around 20 neV.

Another possibility is to use a source that lies at the center of a well studied galaxy cluster. In that case, the magnetic field properties are derived observationally. This is done by studying the Faraday rotation maps of the polarized radio emission from the cluster (see~\cite{Carilli:2001hj} for a review). These studies allow in principle a determination of the full turbulence power  spectrum, yielding the intensity of the magnetic field, its coherence scale and the slope of the turbulence spectrum. A very well studied cluster is Hydra, for which a strong X-ray source is present at the center (Hydra A)~\cite{McNamara:2000kj} . In~\cite{Wouters:2013hua}, X-ray data from the Chandra satellite are analyzed in order to derive constraints on ALP parameters. In the case of X-rays from Hydra, the diagonal terms in the matrix of Eq.~\ref{eq:33mat} can be simplified. Indeed the pair production related opacity is irrelevant in the case of X-rays (so $\tau=0$), and the trace of the matrix is dominated by the effective photon mass for $m_a\lesssim 10^{-11}\;\rm eV$. So the constraints are expected to extend to arbitrarily low ALP masses below that value. In~\cite{Wouters:2013hua}, the irregularity is estimated by performing $\chi^2$ tests when deriving the energy spectrum with a forward folding method. ALP parameters yielding a too high level of irregularity compared to the data are excluded. The corresponding exclusion curve is displayed in Fig.~\ref{fig:excl_x}. It turns out this analysis improves the previous constraints in that mass range from the non-observation of gamma-rays associated with SN 1987 A~\cite{Brockway:1996yr}.

\section{Outlook}\vspace{.5cm}

The study of the high-energy Universe is potentially a nice way to search for axion-like particles. The problem of the transparency of the Universe to gamma-rays can provide a interesting observable, that requires nevertheless the observation of a large number of TeV sources to be robust. This can be achieved with the next generation of Cherenkov telescopes such as CTA. The photon/ALP mixing in astrophysical sources is intrinsically a stochastic process. That fact makes difficult the use of the transparency observations to derive constraints on the ALP parameters. It is noted however that the turbulence of the astrophysical magnetic fields has the effect of inducing irregularities in the energy spectra of sources. The statistical properties of the induced irregularity can be predicted and are used to set limits on the ALP coupling to photons. Because the method is insensitive to the polarization, these constraints go beyond classic ALPs and apply to both $F\tilde{F}$ and $F^2$ types of couplings. Two examples of limits are given in the case of TeV and X-ray observations of high-energy emitting sources inside clusters of galaxies.

\vspace{2cm}

\bibliography{brun}

\end{document}